\documentclass[graybox]{svmult}

\usepackage{type1cm}        
\usepackage{makeidx}         
\usepackage{graphicx}        
\usepackage{multicol}        
\usepackage[bottom]{footmisc}

\usepackage{newtxtext}       %
\usepackage{newtxmath}       

\makeindex             

\begin{document}

\title*{Time-Distance Helioseismology of Deep Meridional Circulation}
\author{S.P. Rajaguru and H.M. Antia}
\institute{S.P. Rajaguru \at Indian Institute of Astrophysics, Koramangala II Block, Bangalore, \email{rajaguru@iap.res.in}
\and H.M. Antia \at Tata Institute of Fundamental Research, Homi Bhabha Road, Colaba, Mumbai \email{antia@tifr.res.in}}
%
%
\maketitle

\abstract*{A key component of solar interior dynamics is the meridional circulation (MC), whose poleward component
in the surface layers has been well observed. Time-distance helioseismic studies of the deep structure of MC, 
however, have yielded conflicting inferences. Here, following a summary of existing results we show how a large center-to-limb
systematics (CLS) in the measured travel times of acoustic waves affect the inferences through an analysis of 
frequency dependence of CLS, using data from the Helioseismic and Doppler Imager (HMI) onboard Solar Dynamics 
Observatory (SDO). Our results point to the residual systematics in travel times as a major cause of differing inferences 
on the deep structure of MC.}

\abstract{A key component of solar interior dynamics is the meridional circulation (MC), whose poleward component
in the surface layers has been well observed. Time-distance helioseismic studies of the deep structure of MC,
however, have yielded conflicting inferences. Here, following a summary of existing results we show how a large center-to-limb
systematics (CLS) in the measured travel times of acoustic waves affect the inferences through an analysis of 
frequency dependence of CLS, using data from the Helioseismic and Doppler Imager (HMI) onboard Solar Dynamics
Observatory (SDO). Our results point to the residual systematics in travel times as a major cause of differing inferences
on the deep structure of MC.}

\section{Introduction}
\label{sec:1}
Large-scale organisation of plasma flows in the convection zones of the Sun and sun-like stars is central 
to a host of problems related to stellar interior dynamics and magnetic dynamos. Poleward meridional flow,
well observed on the solar surface through a variety of techniques, is recognised as a surface component of
deep meridional circulation (MC), which traces back to a nearly century old prediction \cite{eddington25}. There have been a 
good number of theoretical studies of MC with current numerical approaches recognising well that its understanding 
requires solving the complex fluid dynamical problem involving exchanges of energy 
and momentum between convection, rotation, thermal stratification and magnetic fields (see \cite{feathmiesch15} and
references therein). 
Clearly, reliable helioseismic inferences of the deep structure of MC is crucial to 
make progress in this field \cite{toomremjt15}. Recently, helioseismology, especially time-distance helioseismology, has made 
significant progress in this direction, however with unsettling 
differences between the published results \cite{zhaoetal13,jackiewiczetal15,rajaguruantia15,chenandzhao17,
mandaletal18}. A large part of these differences are thought to be 
related to the identification and accounting of a large systematics in travel-time measurements \cite{zhaoetal12}. Here, in this article, 
we summarise these developments and show that further progress in this field depends heavily on understanding the 
origin of this systematics, fully characterising it and removing it reliably.

\section{Time-distance Helioseismology of Deep MC: Current Results}
\label{sec:2}
Most of the inferences on the deep structure of MC have largely been from time-distance helioseismology \cite{duvalletal93}. Travel times
of acoustic waves propagating in meridional planes are measured in deep-focus geometry for a range of travel distances 
$\Delta$ covering depths from the surface down to the base of the convection zone to capture the meridional flows and inverted, 
commonly, in ray theory approximation. Although different studies have followed the above basic method, we refer readers to respective
publications for finer details of the measurement and inversion procedures adopted. We point out that Rajaguru \& Antia (2015)
implemented an in-built mass conservation constraint in terms of the stream function to invert travel times thereby determining both
the meridional ($u_{\theta}$) and radial ($u_r$) components of the flow. The data used by different authors are as follows:
\cite{zhaoetal13}, \cite{rajaguruantia15} and \cite{chenandzhao17} used first two, four and six years of SDO/HMI Doppler data
respectively, \cite{jackiewiczetal15} used two years (2010 - 2012) of Global Oscillation Network Group (GONG) data, 
and \cite{linchou18} have used SOHO/MDI data. 
\begin{figure}[bt]
\includegraphics[scale=.47]{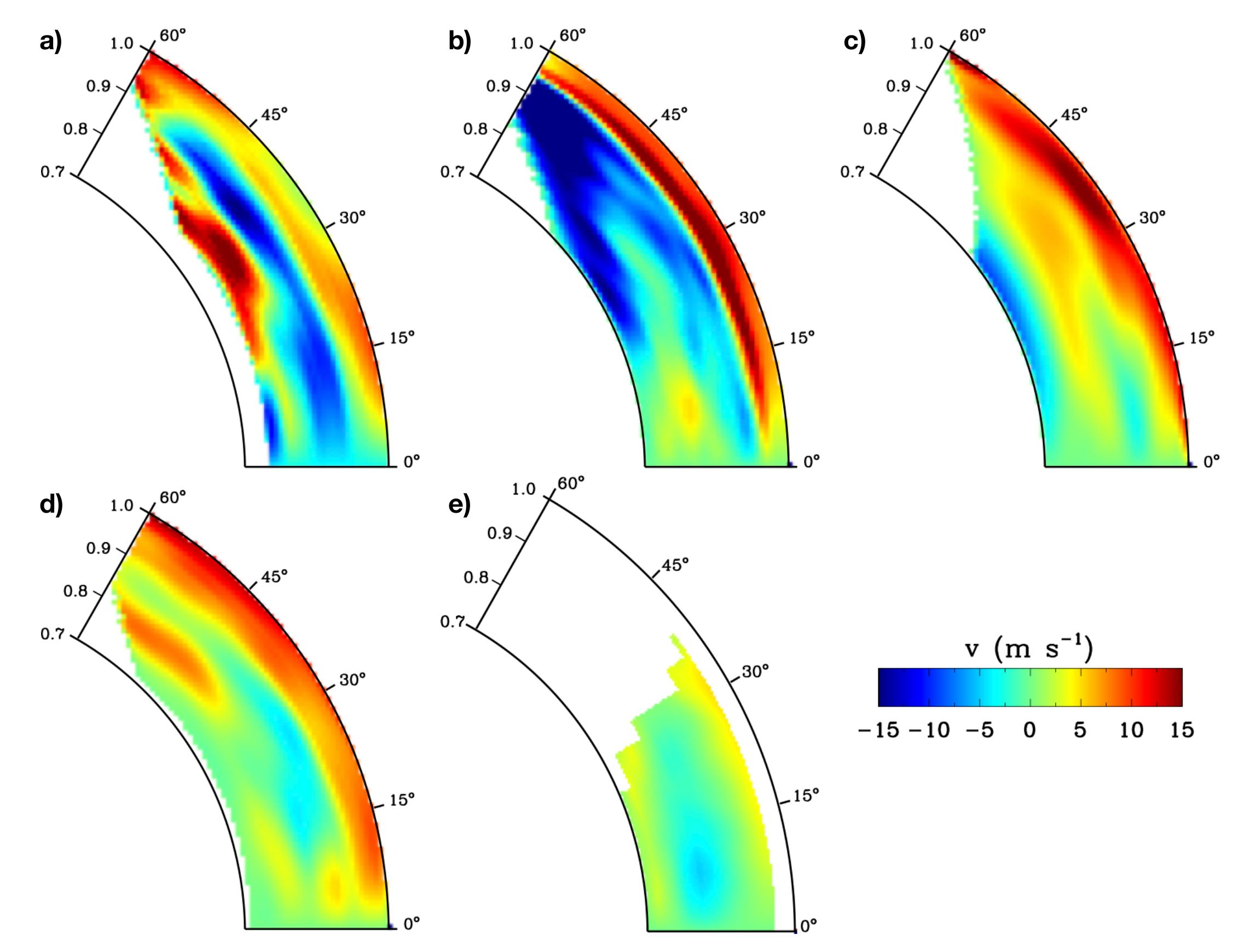}
\caption{Hemispherically symmetrized meridional flow profiles obtained from time-distance helioseismic inversions by
(a) Zhao et al. (2013), (b) Jackiewicz et al. (2015), (c) Rajaguru \& Antia (2015), (d) Chen \& Zhao (2017), and (e) Lin \& Chou (2018).
Poleward flows are postive and equatorward flows are negative. Figure courtesy of Junwei Zhao.}
\label{fig:1}       
\end{figure}

All the different results from the above studies are summarised in Figure 1. Here, meridional flow 
profiles obtained by different authors have been hemispherically symmetrized and plotted.
A prominent feature in the results of \cite{zhaoetal13} and \cite{chenandzhao17} is a 
double-cell MC in depth covering most of the latitudes with the outer cell having a rather shallow return flow at about 0.9R$_{\odot}$. 
Results of \cite{jackiewiczetal15} agree with this shallow return flow but fail to reproduce the deeper second cell of MC.  
Distinct from the above inferences is a single cell deep MC derived by \cite{rajaguruantia15}, with a depth of $\approx$
0.77R$_{\odot}$ for a large-scale reversal of flow. The results of \cite{linchou18} cover only the low latitudes ($<30^{\circ}$) and
show a double-cell structure. It is also worthwhile to note that all the profiles in Figure 1 indicate a
complicated flow structure at low latitudes ($<20^{\circ}$) with two or more reversals of flows over depth, corresponding to
two or more cells.

The above differences among the different studies, employing basically the same method and in some cases even the same data set, 
obviously demand a thorough relook at the analysis procedures, signals and systematics in the measurement. 
The major difference pertaining to the large-scale flow profile, viz.  
single- or double-cell profiles of MC, may, at a basic level, be taken as due to the implementation or not of  
mass conservation constraint in the inversion. However, inaccuracies in the identification and separation of signals from systematics 
could have large impacts in the inferences and we focus on these in the rest of this artcile.

\section{Systematics and Signals in Travel Times}
A major development that led to the above presented time-distance helioseismic studies of MC, in the first place, has been the 
identification and removal of a large systematic center-to-limb effect in the measurements \cite{zhaoetal12,zhaoetal13}, 
especially improving the identification of flow signals in the deeper layers \cite{zhaoetal13,rajaguruantia15,jackiewiczetal15}.
As shown by \cite{zhaoetal12}, the above effect is a large systematic increase in travel time differences against angular distance from the
solar disk center mimicking a radial outflow from the centre towards the limb, and which increases as $\Delta$ increases. This large
center-to-limb systematics (CLS) in travel times is still of unknown origin, although the analyses of \cite{zhaoetal12,zhaoetal13}
involving comparisons of travel times from different observables (corresponding to different heights of formation in the solar
atmosphere) from the SDO/HMI as well as with that from another instrument (SOHO/MDI) pointed to possible physical causes in the solar
atmosphere related to observation height differences. A study by \cite{baldnerschou12} showed that the near-surface granular
convection could affect the wave-propagation in the observable layers leading to a similar effect as the CLS in travel times. 
An empirical correction procedure was suggested  by \cite{zhaoetal12}: estimate CLS from waves traveling in W-E direction 
over an equatorial belt as a function of center-to-limb distance ({\it i.e. longitude}) and subtract it from N-S travel times 
measured against latitude. This prescription was tested using different observables and also validated by comparing with 
direct surface measurements of meridional flows by \cite{zhaoetal12}.
\subsection{Wave-Frequency Dependence of Systematics and Signals}
Given that we do not understand the origin of CLS, hence lack a model to remove it from the measurements 
reliably it is imperative that we devise further diagnostics to characterise it better. In this regard, it has
been suggested \cite{issi17,chenzhao18} that the dependence of CLS on wave frequencies is examined: filter acoustic waves forming resonant p modes 
in frequency domain over several narrow frequency bands, and measure travel times against these frequency bands.
Here, we implement the above by fitting frequency filtered cross-correlations: a total of 15 different frequencies spaced 
every 0.2 mHz interval starting from 2.0 mHz with a FWHM of 1 mHz. Both the N-S and W-E (CLS) cross-correlations are frequency 
filtered, and the meridional flow travel times at each frequency are estimated by subtracting the CLS from the N-S travel times. 
We have used seven years (2010 - 2017) of SDO/HMI data in this study.
\begin{figure}[ht]
\includegraphics[scale=.35]{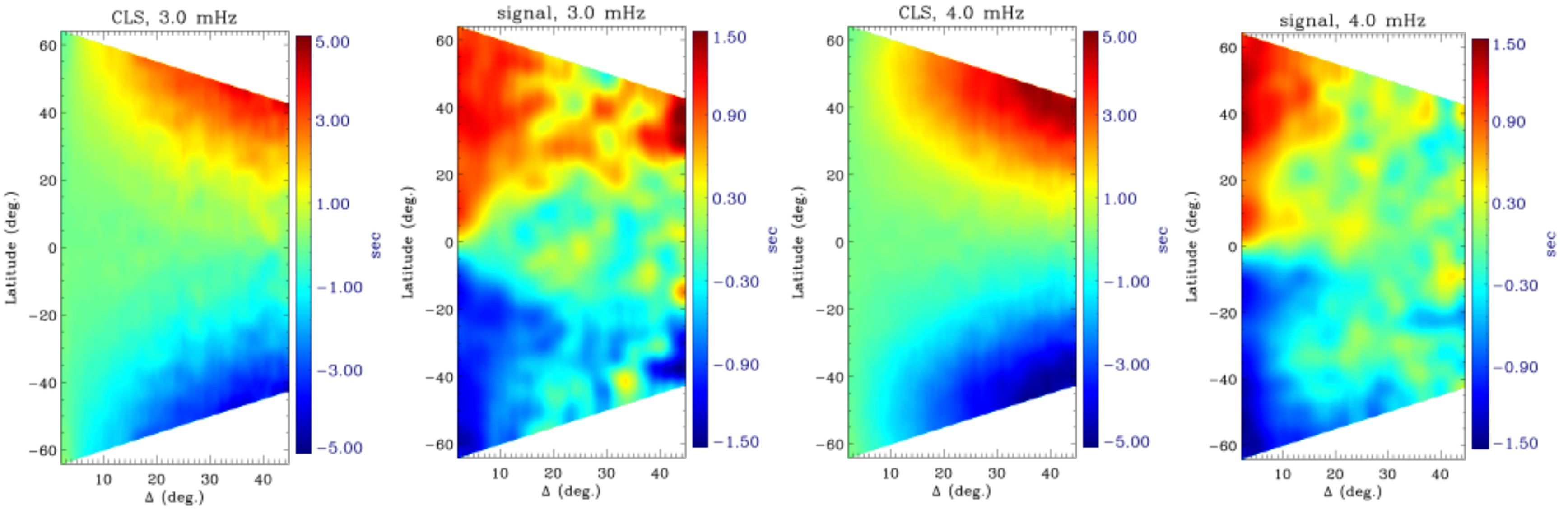}
\caption{Estimates of CLS and meridional flow signals at two representative frequencies, $3.0$ and $4.0$ mHz.
Flow signals are obtained after subtracting the CLS from the N-S travel times at the same frequencies.}
\label{fig:2}       
\end{figure}
Figure 2 displays the CLS as estimated from W-E travel times and the meridional flow signals, which are
obtained after subtracting the CLS from the N-S travel times, for two representative frequencies, $3.0$ and $4.0$ mHz.
The CLS has a strong frequency dependence with rapid increase up to about 4 mHz before levelling off and decreasing slightly 
at higher frequencies. A detailed look at the frequency dependence of CLS and travel times is deferred to a separate publication,
except showing the main impact of this on flow inferences (see next Section). We note that \cite{chenzhao18} have also 
performed a detailed analysis of frequency dependence of CLS via a direct Fourier phase analysis of cross-correlations. 
Since the magnitudes of CLS is several times that of signals due to meridional flows, any slight inaccuracies in the estimation
of CLS would cause significant errors in the signals due to flows. Comparison of signals and CLS in Figure 2 already indicates
a correlation between them, viz. a larger depth gradient of signals correlate with the increased CLS against frequency. It is unclear
as to at which frequency the estimation of CLS is robust and hence the signals too.
\section{Differing Meridional Flow Solutions Over Wave-Frequency}
\label{subsec:2}
The travel times at different frequency bands estimated as above are inverted in ray theory approximation using the same method as 
in \cite{rajaguruantia15} with proper accounting of the frequency dependent ray paths, and the results are shown in Figure 3.
Surprisingly, the solutions differ, with the lower frequency ($3.0$ mHz) travel times yielding a single-cell profile while those 
from the higher frequency ($4.0$ mHz), where the CLS is maximum, yielding a double-cell profile. This property of solutions 
obtained clearly implicate the impact of the strong frequency dependence of CLS and hence points to the role of any residual CLS, 
depending on the accuracy of the procedure to estimate and remove it, in introducing artefacts in the inverted solutions.
\begin{figure}[ht]
\includegraphics[scale=.4]{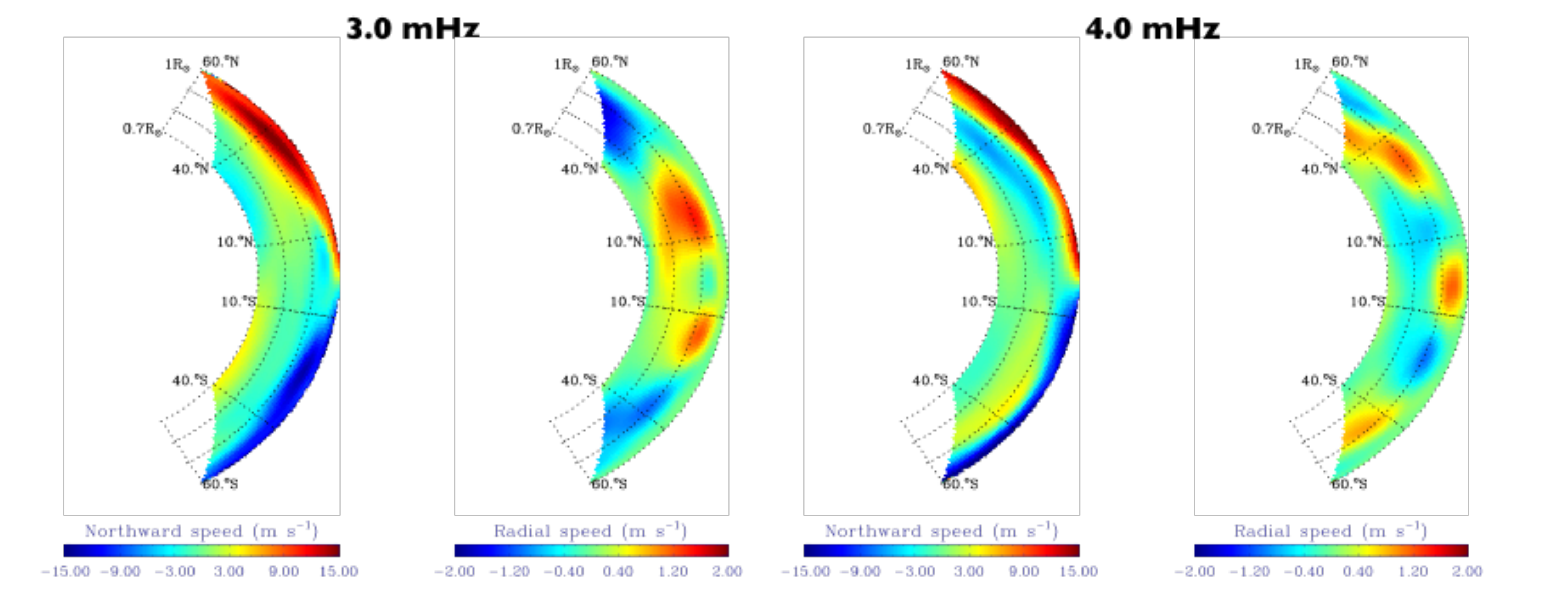}
\caption{Meridional circulation profiles obtained from ray-theory inversion of travel time signals shown in Figure 2.
While travel times measured around $3.0$ mHz give a single-cell profile while those around $4.0$ mHz,
where the CLS is maximum, yield a double-cell profile.}
\label{fig:3}       
\end{figure}
On the other hand if we assume that the frequency dependence of travel times seen in Figure 2 is due to the 
finite wavelength effects and not due to leakage of CLS, then it cannot be modeled by ray-theory approximation
and require wave modeling such as first Born approximation. We defer a detailed analysis of the nature
of solutions against frequency, comparisons with Born approximation treatments of wave-propagation, including the temporal
evolution of MC solutions, to a forthcoming publication. 

\section{Conclusions}
The difficult proposition of measuring the deep MC, down to the base of the convection zone, is further compounded
by the large systematics, CLS,  described and analysed in this work. We have shown that the very different inferences drawn
by different studies in the field are attributable to the CLS: travel time signals estimated
at frequency bands ($\approx$ 4 mHz and above) where the CLS is large yield double-cell MC, while those at lower
frequency bands it is smaller yield single-cell MC. We however point out that we have not established the robustness 
of our estimates of CLS and that it is unclear as to at which frequency the estimation of CLS and the signals are robust and 
hence which of the obtained solutions for deep MC are closer to reality. We therefore caution that more detailed analyses are necessary
to draw firmer conclusions on the above.

\begin{acknowledgement}
The authors are greatly appreciative of their long professional association with Michael Thompson. The first author SPR 
is much grateful to Micheal Thompson for research mentorship, which initiated him in the field of time-distance helioseismology.
Credits are due to Junwei Zhao for providing a compilation in Figure 1.
\end{acknowledgement}
%
%
%

\end{document}